\documentclass[12pt,a4paper]{article}
\usepackage{cite}
\usepackage{amsmath,amssymb,amsfonts}
\usepackage{algorithmic}
\usepackage{graphicx}
\usepackage{textcomp}
\usepackage{mathtools}
\usepackage{caption}
\usepackage{float}
\usepackage{subcaption}
\usepackage[a4paper,left=2.7cm,right=2.7cm,top=2.7cm,bottom=2.7cm]{geometry}
\usepackage{multirow}
\usepackage{array}
\usepackage{setspace}
\usepackage{titlesec}

\titleformat{\section}[block]{\bfseries \large \filcenter}{\Roman{section}.}{1em}{}
\titleformat{\subsection}[block]{\bfseries }{\Alph{subsection}.}{1em}{}

\date{}
\def\BibTeX{{\rm B\kern-.05em{\sc i\kern-.025em b}\kern-.08em
		
		T\kern-.1667em\lower.7ex\hbox{E}\kern-.125emX}}

\title{ Optimal Time-Domain Pulse Width Modulation for Three-Phase Inverters}
\author{Siddharth Tyagi and Isaak Mayergoyz}
\date{%
	\textit{Electrical and Computer Engineering, \\
	 University of Maryland, College Park, MD, 20740, USA}
}

\begin{document}
	
	\setlength{\abovedisplayskip}{8pt}
	\setlength{\belowdisplayskip}{8pt}
	
	\maketitle
	
	\begin{abstract}
			 A novel optimal time-domain technique for  pulse-width modulation  (PWM) in three-phase inverters is presented. This technique is based on the time-domain per phase analysis of three-phase inverters. The role of symmetries on the structure of three-phase PWM inverter voltages and their harmonic contents are discussed. Numerical results highlighting  improvements in the harmonic performance of three-phase inverters are presented. 
	\end{abstract}

	\section{INTRODUCTION}
	
	Inverters are power electronics circuits which convert  DC input voltages into sinusoidal AC output voltages and currents of desired and controllable frequencies. They are widely used in many applications, such as speed control of AC motors in variable-frequency drives (VFDs) \cite{tb1,tb2,tb3}, uninterruptible power supplies (UPS) \cite{ups1,ups2} and for the integration of renewable energy sources with the grid \cite{ren1,ren2,ren3}. Pulse width modulation (PWM) is mainly used to generate output AC voltages in inverters \cite{tb1,tb2,tb3,sp1}. The principle of PWM is to generate inverter voltages which are trains of rectangular pulses. The widths of these pulses are properly modulated to suppress lower-order voltage harmonics at the expense of higher order harmonics. Then, the higher order harmonics are suppressed in output currents and voltages by inductors in the inverter circuit.  
	
	Usually, the H-bridge topology shown in Fig. \ref{3pinv} is used in the design of three-phase inverters. There are a number of ways to generate pulse width modulated voltages and currents for the three-phase circuit inverter shown in Fig. \ref{3pinv}. Space Vector PWM (SVPWM) is the most commonly used method to generate PWM pulses for such inverters \cite{tb1,svp1}.  Over the years, extensive research has been performed on optimal PWM in the frequency domain, that improves upon the SVPWM technique \cite{svp2}-\cite{op6}. In parallel, PWM has been used to selectively eliminate lower-order harmonics. This technique is usually termed as Selective Harmonic Elimination (SHE) and has been extensively studied in the literature \cite{she1,she2,she3,she4,she5,she6,she7,she8}. Other optimal PWM methods based in different ideas have been explored as well \cite{op7,op8,op9}.
	
	The above referenced publications on PWM are concerned with the frequency harmonic content of  line-voltages in three phase inverters. In contrast,  the emphasis in this paper is on the harmonic contents of phase currents or voltages across the resistors shown in Fig. \ref{3pinv}.  This is achieved by using the time-domain per-phase analysis to find analytical  expressions for phase currents, which are then used for minimization of their harmonic-contents. Furthermore, in the framework of the developed technique, specific lower-order harmonics can be completely eliminated by imposing certain constraints on the minimization problem. In this way, SHE is achieved simultaenously  with minimization of Total Harmonic Distortion (THD). 
	
	This manuscript is organized as follows. In Section II, we present a time-domain analysis of PWM for three-phase inverters. The PWM voltages are fully characterized by switching time-instants. The exact analytical solutions for currents and voltages in the circuit shown in Fig. \ref{3pinv} can be obtained in terms of these time-instants. By using these analytical solutions, the problem of optimal PWM design can be framed as a minimization problem. 
	
	\indent Symmetries play an important role in the formation of PWM line-voltages in three-phase inverters. In Section III, we discuss the mathematical and physical aspects of symmetries involved in the performance of PWM inverters. It turns out that these symmetries considerations impose specific constraints on the switching time-instants that describe PWM three-phase line-voltages. In Sections IV, some mathematical details of the optimization technique are discussed, and numerical results are presented. 
	
	\section{TIME-DOMAIN ANALYSIS}
	
	\subsection{Per-Phase Analysis of Three-Phase Inverter}
	
\begin{figure}[h]
		\centering
		\includegraphics[width=4in]{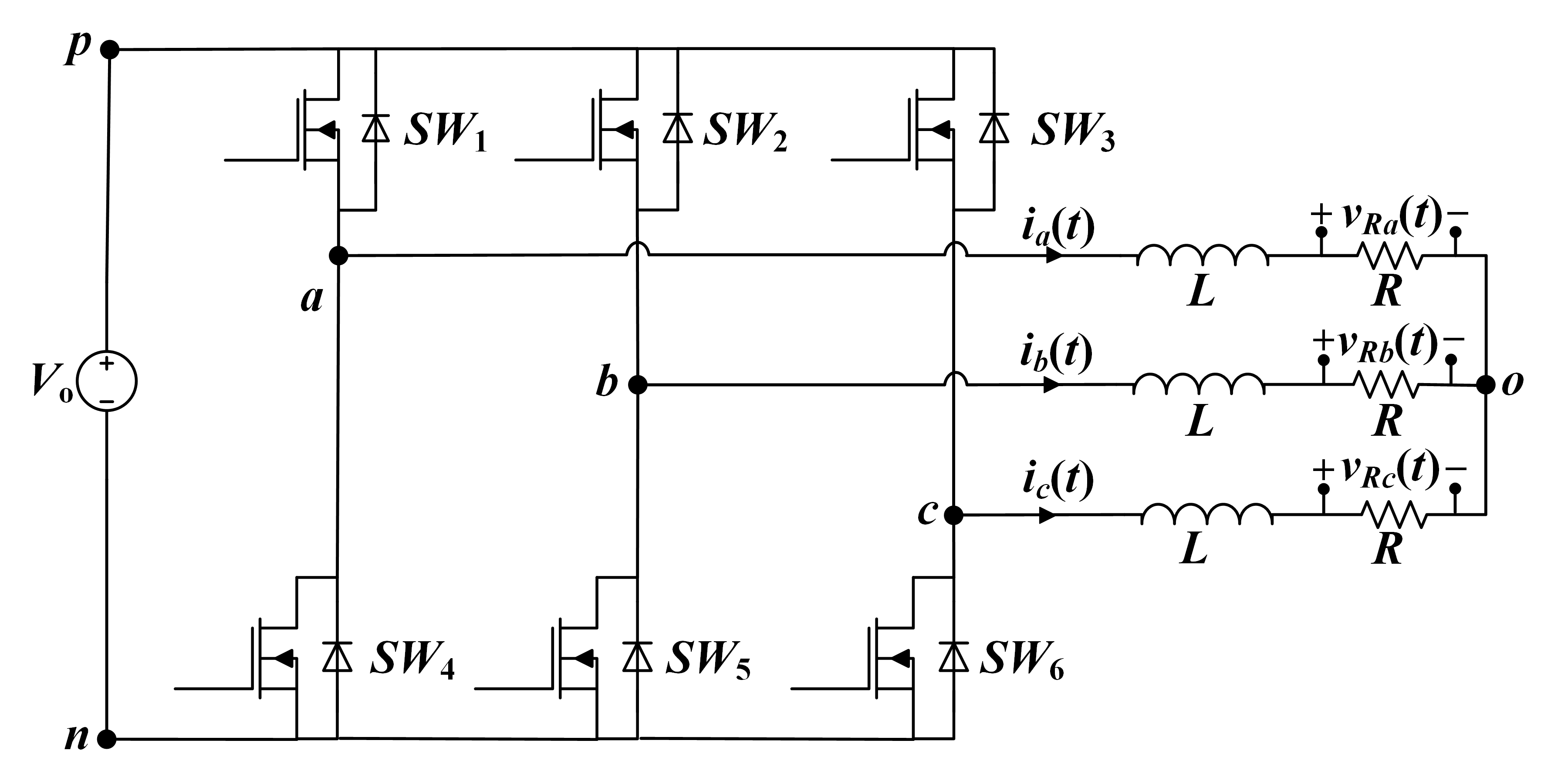}
		\caption{3-phase H-Bridge Inverter}
		
		\label{3pinv}
	\end{figure}

	A three-phase H-bridge inverter is shown in Fig. \ref{3pinv}. Here,  $V_0$ is the DC input voltage. The 3-phase Y-type load connected to the inverter is modeled by using inductors (\textit{L}) and resistors (\textit{R}). It is assumed that the load is balanced. 
	
	We now proceed to derive  per-phase equations for the inverter. Using KVL, we find from Fig. \ref{3pinv} that the phase-currents  $i_a(t)$ , $i_b(t)$ and $i_c(t)$ satisfy the following equations: 
   \begin{equation}     L\frac{di_a}{dt}(t)+ Ri_a(t) =  v_{a}(t) - v_{o}(t),  \label{kvl1} \end{equation}
   \begin{equation}     L\frac{di_b}{dt}(t)+ Ri_b(t) =  v_{b}(t) - v_{o}(t),  \label{kvl2} \end{equation}
   \begin{equation}     L\frac{di_c}{dt}(t)+ Ri_c(t) =  v_{c}(t) - v_{o}(t),  \label{kvl3} \end{equation}
	where $v_a(t), v_b(t), v_c(t)$ and $v_o(t)$ are potentials of nodes $a, b, c$ and $o$, respectively, measured with respect to some reference node, for instance, node $n$. 
 Similarly, using KCL, we find
	\begin{equation}i_a(t) + i_b(t) + i_c(t)= 0.\label{kcl} \end{equation}
	Adding equations (\ref{kvl1}), (\ref{kvl2}) and (\ref{kvl3}) and using (\ref{kcl}), we get the following expression for $v_o(t)$:
	 	\begin{equation} v_{o}(t) = \frac{v_{a}(t) + v_{b}(t) + v_{c}(t)} {3}. \label{von} \end{equation}
	 By substituting the last equation into equation (\ref{kvl1}), we find:
	\begin{equation}
	\begin{split}
	L\frac{di_a}{dt}(t) &+ Ri_a(t) =  v_{a}(t) -  \frac{v_{a}(t) + v_{b}(t) + v_{c}(t)} {3} \\
	             &= \frac{2}{3}v_{a}(t) - \frac{1}{3} \big[v_{b}(t) + v_{c}(t)\big] \\
	             &= \frac{1}{3} \big[v_{a}(t) - v_{b}(t) + v_{a}(t) - v_c(t)\big].
	\end{split}
	\end{equation}
	Thus, 
	\begin{equation}
	L\frac{di_a}{dt}(t)+ Ri_a(t) = \frac{1}{3} \big[v_{ab}(t) + v_{ac}(t) \big]. \label{pha}
	\end{equation}
	Similarly, we can derive:
	\begin{equation}
	L\frac{di_b}{dt}(t)+ Ri_b(t) = \frac{1}{3} \big[v_{bc}(t) + v_{ba}(t) \big] , \label{phb}
	\end{equation}
	\begin{equation}
	L\frac{di_c}{dt}(t)+ Ri_c(t) = \frac{1}{3} \big[v_{ca}(t) + v_{cb}(t) \big] . \label{phc}
	\end{equation}
	We note here that the right-hand sides of equations (\ref{pha}), (\ref{phb}) and (\ref{phc}) depend on the PWM line-voltages which are generated by the inverter, while the left-hand sides of the above equations contain the phase-currents. Thus, (\ref{pha}), (\ref{phb}) and (\ref{phc}) can be interpreted as the per-phase model of the inverter. These equations  shall be used to derive the analytical solution of the phase-currents, and  optimize the harmonic-performance of the inverter. 
	It is instructive to highlight the following similarity of the above time-domain per-phase model of the inverter to the frequency domain per-phase analysis of 3-phase AC circuits under balanced operation. Indeed, once we obtain the analytical solution for the phase current $i_a(t)$ from equation (\ref{pha}) [as discussed in the next section], the analytical expressions for $i_b(t)$ and $i_c(t)$ are easily obtained as  versions of $i_a(t)$ time-shifted by $\frac{T}{3}$ and $\frac{2T}{3}$, respectively. This is the essence of per-phase analysis, where solution for currents and voltages in one phase yields the complete information about currents and voltages in other phases by appropriate time-shifts. 
	
	It is interesting to point out that the right-hand sides of per-phase equations (\ref{pha})-(\ref{phc}) are two-level voltages produced by single-level pulse width modulated line-voltages $v_{ab}(t), v_{bc}(t)$ and $v_{ca}(t)$. This is in clear contrast with single-phase PWM inverters, where the currents are driven by single level line-voltages. 
	  
	\subsection{Analytical Expression for Phase Currents}
\begin{figure}[h]
		\centering
		\includegraphics[width=4in]{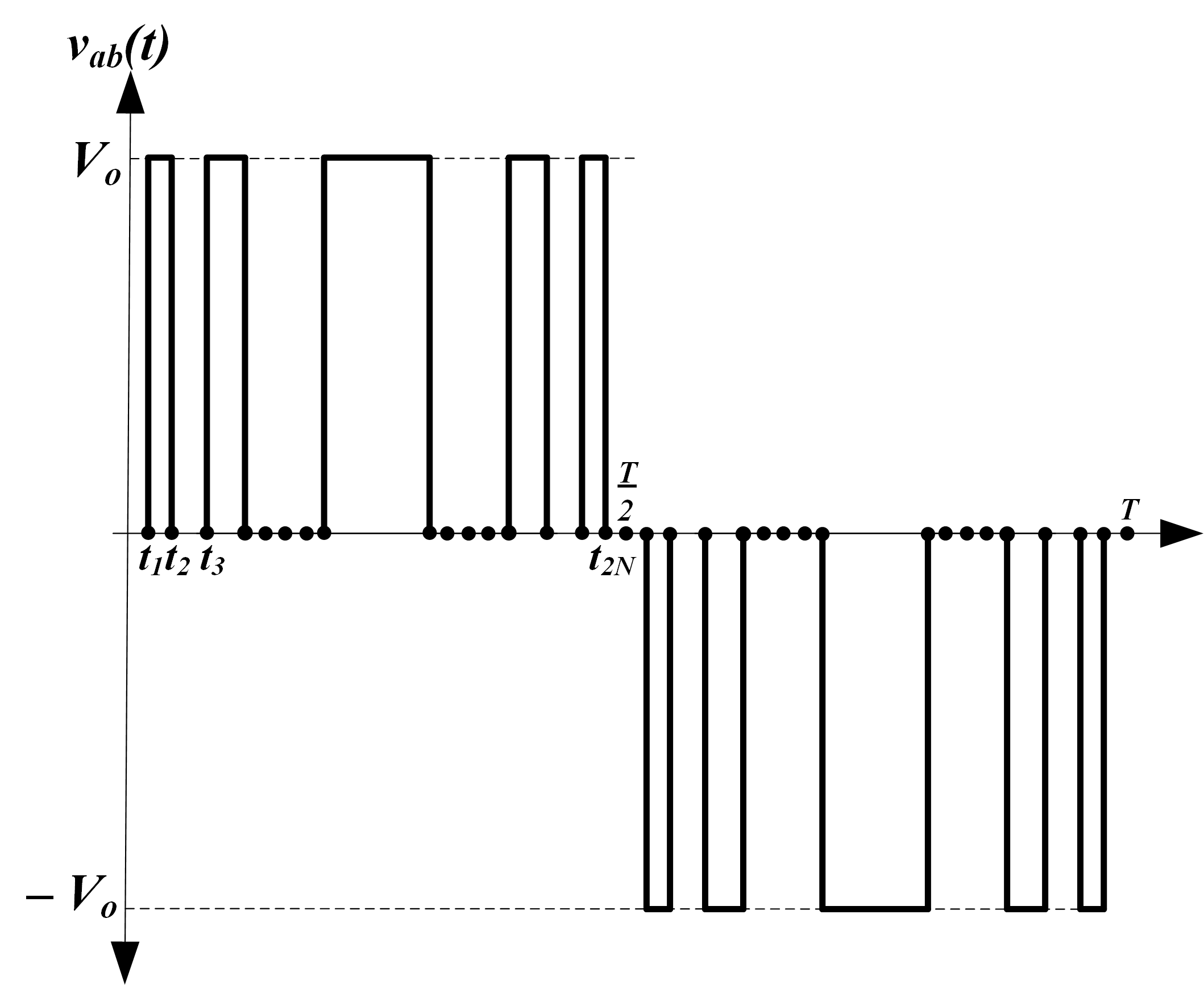}
		\caption{Structure of Output Line-Voltage}
		\label{vabdiag}
	\end{figure}

	For three-phase inverters, the line-voltages are periodic trains of rectangular pulses, as shown in Fig. \ref{vabdiag}. Let  $T$ be the time-period of these voltages, and the associated frequency be $ \omega = \frac {2\pi}{T}$. If the number of pulses in the interval $\big(0, \frac{T}{2}\big)$ is $N$, then these rectangular pulses can be described by a sequence of strictly monotonically increasing switching time-instants $ t_1, t_2, ... , t_{2N} $. 
	
	
	We shall now proceed to derive the expressions for the phase currents $i_a(t), i_b(t)$ and $i_c(t)$ as functions of these switching time-instants. Let $i_{ab}(t)$ and $i_{ac}(t)$ be solutions to the following equations: 
	\begin{equation} L\frac{di_{ab}}{dt}(t) + Ri_{ab}(t) = v_{ab}(t), \label{iabeq} \end{equation}
	\begin{equation} L\frac{di_{ac}}{dt}(t) + Ri_{ac}(t) = v_{ac}(t). \label{iaceq} \end{equation}
	Then, according to the superposition principle for linear differential equations, the solution $i_a(t)$ to equation (\ref{pha}) can be written as: 
	\begin{equation} i_a(t) = \frac{1}{3}\big[ i_{ab}(t) + i_{ac}(t) \big]. \label{iaan} \end{equation} 
	Similar expressions can be written for the other phase currents. 
	
	We begin by solving equation (\ref{iabeq}) for $i_{ab}(t)$ in terms of the switching time-instants that describe the line-voltage $v_{ab}(t)$. Let $v_{ab}(t)$ be a train of rectangular pulses as shown in Fig. \ref{vabdiag}, and hence it can be described by the time-instants $ t_1, t_2,..., t_{2N} $. The voltage $v_{ab}(t)$ must have half-wave symmetry to eliminate even harmonics. This means that:
	\begin{equation} v_{ab}\bigg(t + \frac{T}{2}\bigg) = - v_{ab}(t). \label{hws1}\end{equation}
	Thus, $ v_{ab}(t) $ can be completely characterized by its values in the interval $ 0 \leq t \leq \frac{T}{2} $. 
	
	It is clear that the following formula is valid for $v_{ab}(t)$:
	\begin{equation}
	v_{ab}(t)=\begin{cases}
	0, & \text{if $ t_{2j} < t < t_{2j+1} $},\\
	V_0, & \text{if $ t_{2j+1} < t < t_{2j+2} $},
	\end{cases} \label{vt}
	\end{equation}
	where $ j = 0,1,2,...., N, $  and
	\begin{equation} t_0 = 0, \; \; \;  t_{2N+1} = \frac{T}{2}. \label{tif}\end{equation}
From equations (\ref{iabeq}) and (\ref{vt}), we find that: 
	\begin{equation}
	i_{ab}(t)=\begin{cases}
	A_{2j+1}e^{-\frac{R}{L}t}, & \text{if $ t_{2j} < t < t_{2j+1} $},\\
	A_{2j+2}e^{-\frac{R}{L}t} + \frac{V_0}{R}, & \text{if $ t_{2j+1} < t < t_{2j+2}$},
	\end{cases} \label{it}
	\end{equation}
	where the constants $A_{2j+1}$ and $A_{2j+2}$ must be determined by using the continuity of electric current $i_{ab}(t)$ at times $t_{2j}$ and $t_{2j+1}$ as well as the half-wave symmetry boundary condition:
	\begin{equation} i_{ab}\bigg(\frac{T}{2}\bigg) = - i_{ab}(0), \label{antiper}\end{equation}
	imposed by the half-wave symmetry [see equation (\ref{hws1})] of $v_{ab}(t)$.
	
	From formula (\ref{it}), using the continuity of $i_{ab}(t)$ at the time-instants $ t_1, t_2, ... , t_{2N} $, as well as the boundary condition (\ref{antiper}), we arrive at the following simultaneous equations:
	\begin{align}
	 A_2 - A_1   & = - \frac{V_0}{R}e^{\frac{Rt_1}{L}}, \label{int2} \\
	A_3 - A_2 &=   \frac{V_0}{R}e^{\frac{Rt_2}{L}}, \label{int3} \\
	 &	\vdotswithin{=} \nonumber  \\
	A_{2j} - A_{2j-1} &= -  \frac{V_0}{R}e^{\frac{Rt_{2j-1}}{L}}, \label{int4}\\
    A_{2j+1} - A_{2j} &=  \frac{V_0}{R}e^{\frac{Rt_{2j}}{L}}, \label{int5}\\
	  &  \vdotswithin{=} \nonumber \\
	A_{2N+1} - A_{2N} &=  \frac{V_0}{R}e^{\frac{Rt_{2N}}{L}}, \label{int6}
	\end{align}
	and
	\begin{equation}
		A_{1}+ A_{2N+1}e^{-\frac{RT}{2L}} =  0. \label{int7}
	\end{equation}
	These are linear simultaneous equations with a sparse two-diagonal matrix. These equations can be analytically solved as follows.  
    Adding all the equations from (\ref{int2}) to (\ref{int6}), we find
     \begin{equation} A_{2N+1}- A_{1} =  \frac{V_0}{R}\sum_{j=1}^{2N}(-1)^j e^{\frac{Rt_{j}}{L}}. \label{int8}\end{equation}
   Solving equations (\ref{int7}) and (\ref{int8}), we derive:
	\begin{equation} A_{2N+1} = \frac{V_0}{R} \frac {\sum_{j=1}^{2N}{(-1)^j e^{\frac{R}{L}t_j}}}{1 + e^{-\frac{RT}{2L}}},   \label{A2k1} \end{equation}
	\begin{equation} A_1 = -\frac{V_0}{R} \frac {\sum_{j=1}^{2N}{(-1)^j e^{\frac{R}{L}t_j}}}{1 + e^{-\frac{RT}{2L}}} e^{-\frac{RT}{2L}} . \label{A1} \end{equation}  
	Having found $A_1$ all other $A$-coefficients can be computed using the following formula:
	\begin{equation} 
	A_j = A_1 + \frac{V_0}{R} \sum_{n=1}^{j-1}{(-1)^n e^{\frac{R}{L}t_n }} \; \;  \text{ for } j=2,3,...,2N. \label{Aj} 
	\end{equation}
	The above formula is obtained by adding the first $j-1$ equations (\ref{int2})-(\ref{int6}).
	
	By using the above formulas for the $A$-coefficients in equation (\ref{it}), we find the general analytical solution for the current $i_{ab}(t, t_1, t_2, ... , t_{2N})$ in terms of the switching time-instants that describe the voltage $v_{ab}(t)$.   

	Next, we find the analytical expression for $i_{ac}(t)$.  We observe that, in order to eliminate all harmonics of orders divisible by three in the line-voltages, the following translational-symmetry condition must be satisfied:
	\begin{equation} v_{ab}(t) = v_{bc}\bigg(t + \frac{T}{3} \bigg) = v_{ca}\bigg(t - \frac{T}{3} \bigg) \label{ts1}. \end{equation}
	Furthermore, 
	\begin{equation} 
	\begin{split}
	v_{ba}(t) &= - v_{ab}(t), \nonumber \\   v_{cb}(t) &= - v_{bc}(t), \label{ts2}\\  v_{ac}(t) &= - v_{ca}(t) \nonumber .
	\end{split}
	\end{equation}	
	The above two equations imply that $i_{ac}(t)$ is a time-shifted version of $i_{ab}(t)$. Namely, $i_{ac}(t)$ can be expressed as a function of the switching time-instants  $t_1, t_2, ... , t_{2N}$ as follows:
	\begin{equation}
		i_{ac}(t, t_1, ... , t_{2N}) =- i_{ab}\bigg(t + \frac{T}{3},  t_1,..., t_{2N} \bigg).  \label{iact}
	\end{equation} 
	Substituting the analytical expression for $i_{ab}(t)$  given by equations (\ref{it}) and (\ref{A2k1})-(\ref{Aj})  as well as equation (\ref{iact}) in equation (\ref{iaan}), we arrive at the following expression for $i_a(t)$:
	\begin{equation}
	\begin{split}
	i_a(t) = \frac{1}{3}\bigg[ & i_{ab}(t,t_1, t_2,..., t_{2N}) \nonumber \\ &- i_{ab}\bigg(t + \frac{T}{3}, t_1, t_2,..., t_{2N}\bigg)\bigg].
	\end{split}
	\label{ia} 
	\end{equation}
  	 The latter implies that:
  	 \begin{equation}
  	 i_a(t) = i_a(t, t_1, t_2,..., t_{2N}), \label{iafunc}
  	 \end{equation}
  	 which means that the analytical expression for the phase-current $i_a(t)$ in terms of the switching time-instants $t_1, t_2, ... , t_{2N}$ that describe three-phase line-voltages can be obtained.
  	 
	Before proceeding with the further discussion, we make the following important observation. Equations (\ref{ts1}) and (\ref{ts2}) imply that all three-phase line-voltages can be described by a \textit{single} sequence of strictly monotonically increasing switching time-instants $ t_1, t_2,..., t_{2N} $. However, it turns out that not any given sequence of strictly monotonically increasing time-instants $ t_1, t_2,..., t_{2N} $ may, in general, represent three-phase PWM line-voltages. The reason is that time-symmetries of line voltages, as well as the KVL requirement that the voltages $v_{ab}(t)$, $v_{bc}(t)$ and $v_{ca}(t)$ must add up to zero,  impose specific constraints on the switching time-instants that describe 3-phase PWM line-voltages. Furthermore, there are also constrains imposed by the fact that only two switches in the same leg of the three-phase inverter in Fig. \ref{3pinv} are usually operated simultaneously. The detailed discussion of these constraints is presented in section III.
	
	\subsection{Time-Domain Optimization}
	
	Now, we shall describe the central idea of the optimal time-domain pulse width modulation technique.  
	
	We begin with deriving the expression for the fundamental harmonic component of $i_a(t)$.  The fundamental harmonic components of the line-voltages $v_{ab}(t)$, $v_{bc}(t)$ and $v_{ca}(t)$ can be written as follows:
	 \begin{align}
	  v_{ab,1}(t) &= V_m \sin(\omega t) \label{vab1}, \\
	  v_{bc,1}(t) & = V_m \sin\bigg(\omega t - \frac{2\pi}{3}\bigg), \label{vbc1} \\
	  v_{ca,1}(t) &= V_m \sin\bigg(\omega t + \frac{2\pi}{3}\bigg). \label{vca1}
	  \end{align}
	Using equations (\ref{iabeq}) and (\ref{iaceq}), along with (\ref{ts2}),  (\ref{vab1}) and (\ref{vca1}), the corresponding fundamental harmonic components of  currents $i_{ab}(t)$ and $i_{ac}(t)$ can be expressed as follows:
	 \begin{equation}
	 		i_{ab,1} (t)= \frac{V_m}{\sqrt{R^2 + (\omega L)^2}} \sin(\omega t - \phi) ,
	 \label{iab1}
	 \end{equation}
	 \begin{equation}
	  i_{ac,1} (t)= -\frac{V_m}{\sqrt{R^2 + (\omega L)^2}} \sin\bigg(\omega t  + \frac{2\pi}{3}- \phi \bigg),
	 \label{iac1}
	 \end{equation}
	 where 
	 \begin{equation} \tan \phi = \frac{\omega L}{R}. \label{phi} \end{equation}
	 By using equations (\ref{iab1}) and (\ref{iac1}) as well as formula (\ref{iaan}), the fundamental harmonic component of $i_a(t)$ can be obtained. Namely,
	 \begin{equation}
	i_{a,1}(t)  = \frac{V_m}{\sqrt{3} \sqrt{R^2 + (\omega L)^2}} \sin \bigg(\omega t - \phi - \frac{\pi}{6}\bigg),
	 \end{equation}
	 which can also be written as:
	 \begin{equation}
	   i_{a,1}(t) = I_m \sin \bigg(\omega t - \phi - \frac{\pi}{6}\bigg) ,
	   \label{ia1}
	 	 \end{equation}
     where
	 \begin{equation} V_m = \sqrt{3} I_m \sqrt{R^2+(\omega L)^2}. \label{Vm} \end{equation}
	 
	Next, we want to find the switching time-instants  $ t_1, t_2,..., t_{2N} $ in equation (\ref{iafunc}) by  minimizing in certain sense the difference:
	\begin{equation}
		e(t, t_1, t_2, ... , t_{2N}) =  i_a(t, t_1, t_2,..., t_{2N}) - i_{a,1}(t). \label{e}
	\end{equation}
	 
	  Namely, the optimal time-domain pulse width modulation problem can be stated as follows: find such time-instants $ t_1, t_2,..., t_{2N} $ that the following quantity $E_2$ reaches its minimum value:
	\begin{equation}
	\begin{split}
	 E_2(t_1,..., t_{2N}&)=\int_{0}^{\frac{T}{2}} \bigg[ i_a(t,t_1,..., t_{2N}) \nonumber \\ &-I_m \sin \bigg(\omega t - \phi - \frac{\pi}{6}\bigg)\bigg]^2 dt  \label{E2} 
	\end{split}
	\end{equation}
It is apparent that this is the least squares optimization. In mathematical terms, the latter means the optimization of the error-function $i_a(t) - i_{a,1}(t)$ in the $L_2$-norm.  It is worthwhile to relate the function $E_2(t_1,..., t_{2N})$ to the total harmonic distortion (THD) in phase-current $i_a(t)$. The latter is denoted by $THD_I$, and it is defined as:
	 \begin{equation}
	 	THD_I = \frac{\sqrt{\sum_{n=2}^{\infty}I_n^2}}{I_f^2},
	 \end{equation}
	where $I_f$ is the amplitude of the fundamental harmonic component in $i_a(t)$, while $I_n$ is the amplitude of its  $n^{\text{th}} $ harmonic. It can be easily verified, by substituting $ i_a(t, t_1,..., t_{2N})$ in (\ref{E2}) in terms of its Fourier series expansion and using the orthogonality property of trigonometric functions that the error integral $E_2(t_1,..., t_{2N})$ and $THD_I$ are related by the following equation:
	\begin{equation}
		E_2(t_1,..., t_{2N}) = \big[ (I_f - I_m)^2 + I_f^2 (THD_I)^2  \big] \cdot \frac{T}{4}.
		\label{ethdreln}
	\end{equation}
	The last formula is a special case of the well known Parseval's equality for the Fourier series. It is evident from the above formula that the minimization of the function $E_2(t_1, .., t_{2N})$ leads to a minimization of the THD in the phase-currents. 
	
	It turns out that specific order harmonics in the function $e(t, t_1,..., t_{2N})$ defined in (\ref{e}) can be completely eliminated within the structure of the stated optimization technique. This is done by using constrained optimization. This approach can also be used to ensure that the fundamental harmonic component  $I_f$ of the phase-current has the desired value $I_m$. Namely, the following constraint can be imposed on the switching time-instants that describe the line-voltage $v_{ab}(t)$: 
	\begin{equation}
		\frac{2V_0}{\pi}\sum_{j=1}^{2N}(-1)^{j+1} \cos \omega t_j  = V_m, 
	 \label{fund}
	\end{equation} 
	where $V_m$ is defined by formula (\ref{Vm}). 
	
Similarly, constraints can be imposed to completely eliminate specific order harmonics. For instance, in order to eliminate the $m^\text{th}$ harmonic, the following constraint can be used \cite{tb1,she1,sop1}: 
	\begin{equation}
	\sum_{j=1}^{2N}(-1)^j \cos(m\omega t_j) = 0.
	\label{shecons}
	\end{equation} 
Thus, the optimization technique can be structured to  eliminate specific lower-order harmonics, and minimize the total harmonic content of the remaining higher-order harmonics. It is worthwhile to mention that by using the method of Lagrange multipliers, the stated problem can be reduced to unconstrained optimization. 
	
\section{CONSTRAINTS ON SWITCHING TIMES IMPOSED BY SYMMETRIES}
	\subsection{Symmetries}
	Symmetries play an important role in pulse width modulation of line-voltages in three-phase inverters. Our subsequent discussion deals with the following symmetries. 
	
	\textbf{S1. Translational Symmetry}: The three-phase line-voltages $ v_{ab}(t), v_{bc}(t)$ and $ v_{ca}(t)$ are time-shifted versions of each other. Namely, the following identity is valid:
	\begin{equation} v_{ab}(t) = v_{bc}\bigg(t + \frac{T}{3} \bigg) = v_{ca}\bigg(t - \frac{T}{3} \bigg). \label{S1} \end{equation}
	Translational symmetry ensures that the fundamental harmonic components of the line-voltages form a balanced, positive sequence of three-phase voltages \cite{tb3}. Furthermore, it can be shown that translational symmetry results in the elimination of all harmonics of orders divisible by three.
	
	\textbf{S2. Half-Wave Symmetry}: This symmetry implies that:
	\begin{equation} v_{ab}(t) = - v_{ab}\bigg(t + \frac{T}{2} \bigg). \label{S2} \end{equation}
	The same half-wave symmetry is valid for $v_{bc}(t)$ and $ v_{ca}(t)$.  It can be shown that half-wave symmetry results in the elimination of even-order harmonics in the line-voltages.
	
	\textbf{S3. Quarter-wave symmetry}: The objective of PWM is to generates output voltages that approximate ideal sinusoidal voltages. Hence, it makes intuitive sense to impose the following quarter-wave symmetry condition on the PWM voltages:   
	\begin{equation} v_{ab}(t) =  v_{ab}\bigg( \frac{T}{2} - t \bigg).  \label{S3} \end{equation}
	It is interesting to point out that quarter-wave symmetry (\ref{S3}), half-wave symmetry (\ref{S2}) and periodicity imply that the line-voltage $v_{ab}(t)$ has odd-symmetry. Indeed:
	
	\begin{equation}
	v_{ab}(t) =  v_{ab}\bigg( \frac{T}{2} - t \bigg) = - v_{ab}(T-t) = -v_{ab}(-t).
	\label{oddsymm}
	\end{equation} 

	In addition to the above fundamental symmetry conditions, the three-phase line-voltages must satisfy the following constraints.
		
	\textbf{C1. KVL constraint}: The sum of three-phase line-voltages $v_{ab}(t)$, $v_{bc}(t)$ and $v_{ca}(t)$ equals zero. 
	\begin{equation}   v_{ab}(t) + v_{bc}(t) + v_{ca}(t) = 0 \label{C1}. \end{equation}
	
	\textbf{C2. Switching pattern constraint}: These are constraints related to the fact that only the states of the two switches in the same leg of the inverter can be simultaneously changed. This prevents unnecessary switchings and helps minimize switching-losses \cite{tb1,op1}.
		
	\subsection{Role of symmetries on structure of PWM line-voltages}
	
	\begin{figure*}
		\centering
		\includegraphics[scale=0.9]{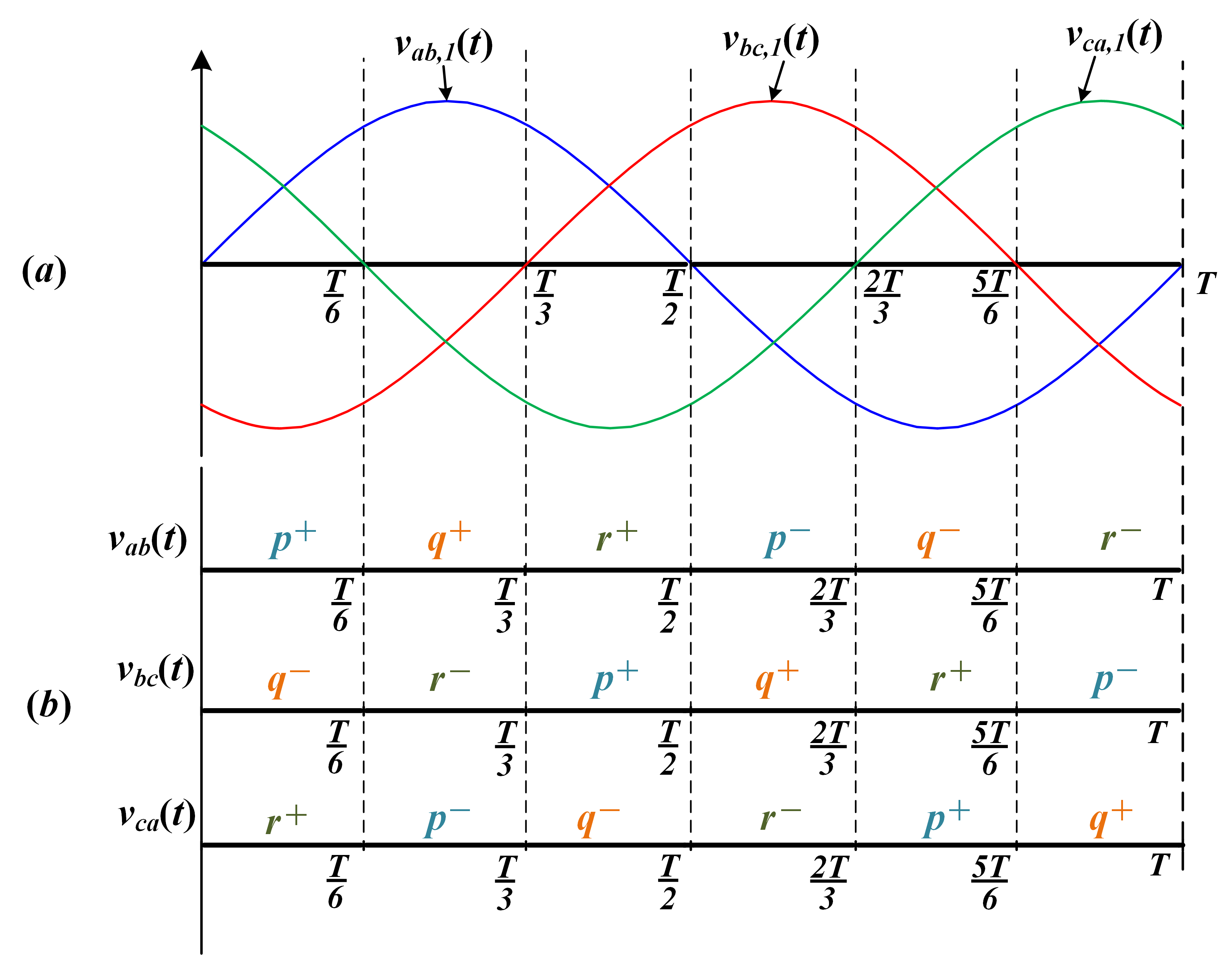}
		\caption{Structure of 3-phase Line-to-line-voltages}
		\label{pqrdiag}
	\end{figure*}
	Next, we discuss the implications of the above symmetries and constraints on the structure of the three-phase PWM line-voltages.
	
	The desired fundamental-components of the line-voltages are shown Fig. \ref{pqrdiag}(a). We begin by dividing the interval $ 0 \leq t \leq T$ into six equal subintervals of length $\frac{T}{6}$. In each of these subintervals, the PWM pulses of the line-voltages can be grouped together to form a \textit{pulse-group}. Thus, for each line-voltage, each subinterval of length $\frac{T}{6}$ can be characterized by a unique pulse-group.       
	
	We first describe the pulse-groups that constitute the PWM line-voltage $v_{ab}(t)$. We label these three pulse-groups in the interval $0 \leq t \leq \frac{T}{2}$ as $p^+, q^+$ and $r^+$, respectively. Since $v_{ab,1}(t)$ is positive in the interval $0 \leq t \leq \frac{T}{2}$,  $v_{ab}(t)$ shall switch between values 0 and $+V_0$ in this interval, as shown in Fig. \ref{vabdiag}. Hence, pulse-groups for $v_{ab}(t)$ in the interval $0 \leq t \leq \frac{T}{2}$ are marked by superscript ``$+$".  Furthermore, as a consequence of half-wave symmetry (\ref{S2}), pulses in the interval  $\frac{T}{2} \leq t \leq T $ are negative copies of the pulses in $0 \leq t \leq \frac{T}{2}$ (see Fig. \ref{vabdiag}). Hence, they can be represented by pulse-groups marked using the labels $p^-, q^-$ and $r^-$, as shown in Fig. \ref{pqrdiag}(b). The pulses that constitute the pulse-groups $p^-, q^-$ and $r^-$ have the same widths but opposite polarities as compared to the corresponding pulses in the $p^+, q^+$ and $r^+$ groups, respectively. Thus, $p^+, q^+$, $r^+$, $p^-, q^-$ and $r^-$ are six distinct pulse-groups which constitute the line-voltage $v_{ab}(t)$. 
	Furthermore, quarter-wave symmetry  (\ref{S3}) for $v_{ab}(t)$ implies the pulses in the $p$ group are mirror images  (with respect to $t =\frac{T}{4}$) of those in the $r$ group.
	
	Next, the translational symmetry (\ref{S1}) can be used to determine the pulse-groups in the six subintervals for the line-voltages $v_{bc}(t)$ and $v_{ca}(t)$. Since these line-voltages are time-shifted versions of $v_{ab}(t)$, the pulse-groups in each of the subintervals for $v_{bc}(t)$ and $v_{ca}(t)$ are as shown in Fig. \ref{pqrdiag}(b). 

	\indent Now, we discuss the implications of the KVL constraint. From Figure 3(b), we observe that in the time interval $0 \leq t \leq \frac{T}{6}$, the line-voltages $v_{ab}(t)$, $v_{bc}(t)$ and $v_{ca}(t)$ have pulses of the $p^+, q^-$ and $r^+$ groups, respectively. Similarly, for the subsequent time intervals of length $\frac{T}{6}$,  the pulse-groups of these three line-voltages are: $(q^+, r^-, p^-), (r^+,p^+,q^-), (p^-,q^+,r^-),$ $(q^-,r^+,p^+)$ and $(r^-,p^-,q^+)$, respectively. It is apparent that for each of these time intervals, two of the line-voltages are represented by pulses from the $p$ and $r$ groups of the same sign, while the other line-voltage pulses belong to the $q$ group of the opposite sign. Thus, KVL equation (\ref{C1}) as well as the translational symmetry implies that for each pulse in the $q^+$ (or $q^-$) group, there are corresponding pulses of the opposite polarity in the $p^-$ (or $p^+$) group and the $r^-$ (or $r^+$) group, such that their total sum is equal to zero. Furthermore, since half-wave symmetry ensures that pulses in the $q^+$ and $q^-$ groups have the same width but opposite signs,  we can arrive at the following important conclusion: half-wave symmetry, translational symmetry and the KVL constraint imply that each pulse in the $q$ group is the sum of two specific pulses of the same sign: one from the $p$ group and one from the $r$ group.

	\subsection{Constraints on Switching Time-Instants}
	
	We now proceed to discuss the constraints that switching time-instants $ t_1, t_2,..., t_{2N} $ must satisfy to represent three-phase PWM line-voltages.
	
	First, we determine the number of pulses in  three-phase PWM line-voltages. Let the number of pulses in the $p$ group be $P$. Because of quarter-wave symmetry, pulses in the $r$ group are mirror images of pulses in the $p$ group of the same sign. For this reason, the number of pulses in the $r$ group also equals $P$.  Let the number of pulses in the $q$ group be $Q$. It is apparent from Fig. \ref{pqrdiag} and KVL that pulses in the $q$ groups must be wider than the pulses in the $p$ and $r$ groups. Furthermore, it was found in the last subsection that pulses in the $q$ group are sums of pulses in the $p$ and $r$ groups of the same sign. This implies that for every pulse in the $q$ group, there must exist one pulse in the $p$ group \textit{and} one pulse in the $r$ group which add to form the given pulse in the $q$ group. This implies that $Q = P$. Thus, we conclude that the number of pulses in the $p$, $q$  and $r$ groups are the same and equal to $P$. This means that $N = 3P = 3Q$. It is desirable that $v_{ab}\big( t= \frac{T}{4} \big) = V_0$ (since $v_{ab,1}(t)$ reaches maximum at $\frac{T}{4}$). For this reason and quarter-wave symmetry, $Q$ is odd. That is, $Q = 2M + 1$, where $M$ is a natural number, and hence $N=3(2M+1)$.
	
	Next, we proceed to obtain the algebraic relations that the switching time-instants $ t_1, t_2,..., t_{6P} $ must satisfy in order to represent three-phase PWM line-voltages. Consider the pulses for line-voltage $v_{ab}(t)$ (see Fig. \ref{vabdiag}) in the interval $ 0 \leq t \leq \frac{T}{6}$, that is the pulses in the $p^+$ group. Each such pulse can be indexed by $l$, where $l = 0,1,2,...,P$. The switching time-instants associated with the $l^\text{th}$ pulse in the $p^+$ group are $t_{2l-1}$ and $t_{2l}$. Clearly, time-instants $t_{2P+2l-1}$ and $t_{2P+2l}$ correspond to the $l^\text{th}$ pulse in the $q^+$ group, while  $t_{4P+2l-1}$ and $t_{4P+2l}$ correspond to the $l^\text{th}$ pulse in the $r^+$ group.
	
	Our discussion in the previous subsection suggests that switching time-instants for pulses is the $q^+$ and $r^+$ groups can be obtained from time-instants $t_{2l-1}$ and $t_{2l}$ in the $p^+$ group. Indeed, since the pulses in $r^+$ group are mirror images of those in the $p^+$ group, the corresponding time-instants for pulses in the $r^+$ group can be obtained from quarter-wave symmetry. Furthermore, as a consequence of half-wave symmetry, translational symmetry and KVL, each pulse in the $q^+$ group is the sum of specific pulses in the $p^+$ and $r^+$ groups , and hence the time-instants for pulses in the $q^+$ group can also be obtained in terms of switching time-instants in the $p^+$ group. We now proceed to derive these relations.

	\begin{figure}[h!]
		\centering
		\includegraphics[width=5in]{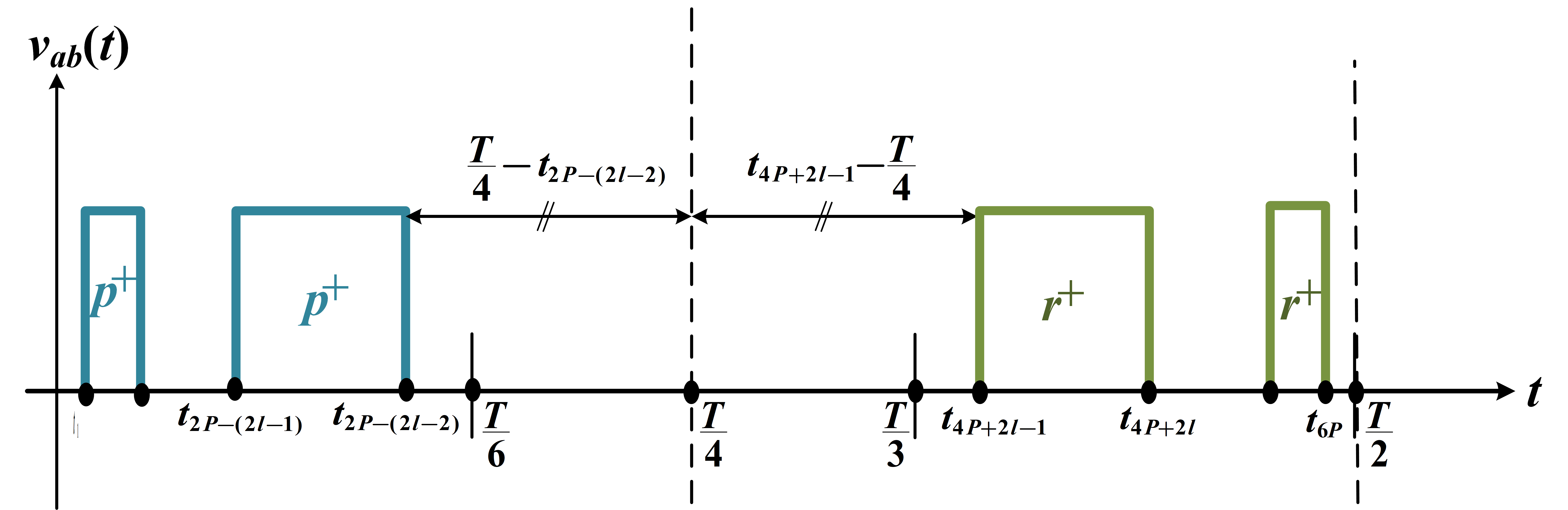}
		\caption{Relation between time-instants in $p$ and $r$ groups}
		\label{prdiag}
	\end{figure}
	
	The algebraic relations between switching time-instants for pulses in the $p^+$ and $r^+$ groups are easily obtained using the quarter-wave symmetry (\ref{S3}), as shown in Fig. \ref{prdiag}. As a consequence of quarter-wave symmetry, for every time-instant defining a rising (falling) edge of a pulse in the $p^+$ group, there is a corresponding time instant defining a falling (rising) edge of a pulse in the $r^+$ group, and these two time-instants are related. Thus, for the $l^\text{th}$ pulse in the $p^+$ group, the corresponding time-instants for pulses the $r^+$ group can be obtained as follows:
	\begin{align}
	t_{4P + 2l-1} &= \frac{T}{2} - t_{2P - (2l-2)}, \label{tpr11} \\
	t_{4P + 2l } &= \frac{T}{2} - t_{2P - (2l-1)}, \label{tpr12}
	\end{align}
	where $t_{2P - (2l-2)}$ and $t_{2P -(2l-1)}$ are time-instants for pulses in $p^+$ group, and  $l = 1,2...,P$. 
	
\begin{figure*}
		\centering
		\includegraphics[width=\linewidth]{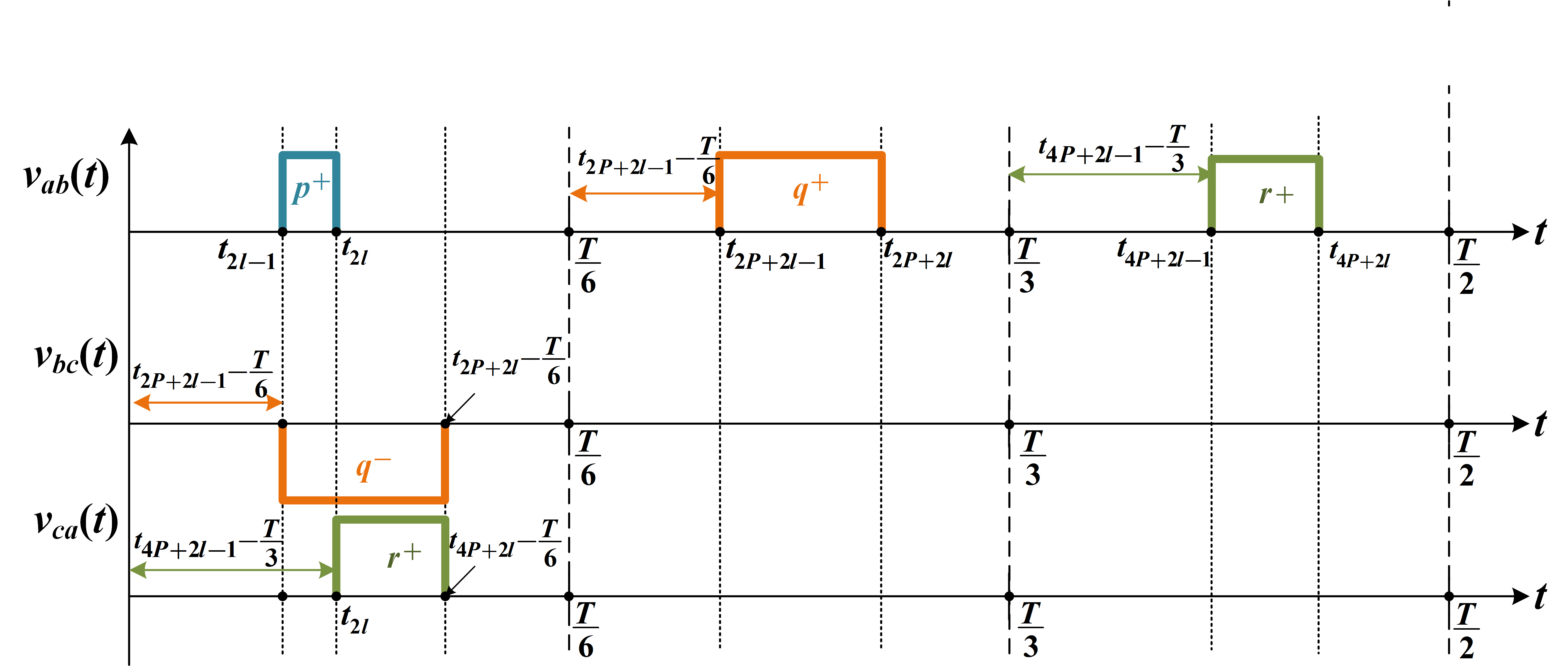}
		\caption{Structure of  pulses when $l$ = odd}
		\label{oddpulse}
	\end{figure*}

\begin{figure}[h!]
		\centering
		\includegraphics[width=3.5in]{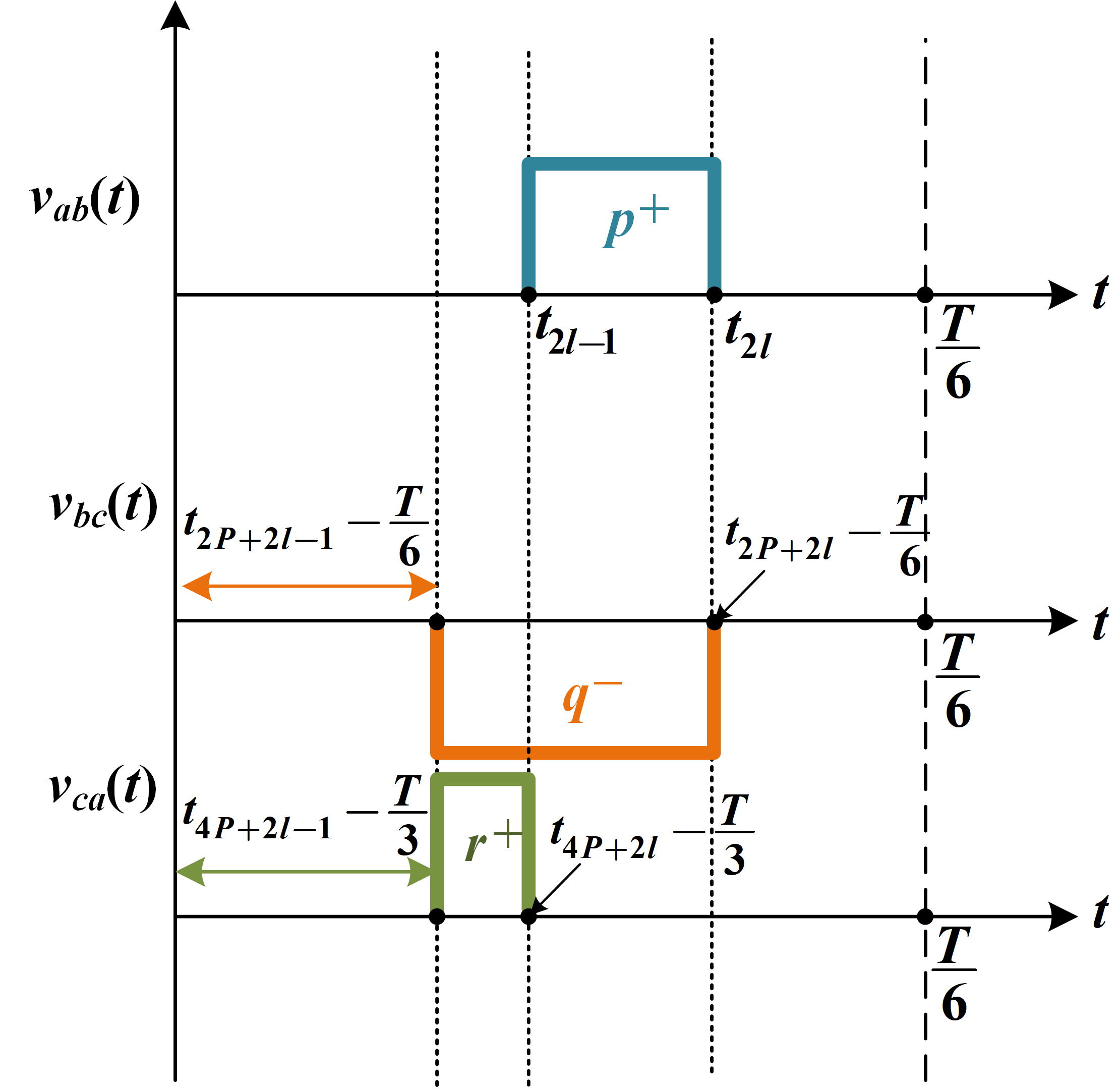}
		\caption{Structure of pulses when $l$ = even}
		\label{evenpulse}
	\end{figure}

	We now proceed to obtain switching time-instants for pulses in the  $q^+$ group in terms of switching time-instants in the $p^+$ group. It can be shown that the single-leg switching constraints (C2) lead to two specific patterns on how the KVL constraint (C1) is realized \cite{tb1,tb2}. Namely, for odd-pulses (i.e. when $l$ is odd), the KVL compensation of the corresponding $p$, $q$ and $r$ pulses occur as shown in Fig. \ref{oddpulse}. Whereas, for even-pulses (i.e. when $l$ is even), this compensation occurs as shown in Fig. \ref{evenpulse}. These figures are used below to derive the formulas for switching time-instants for pulses in the $q^+$ group in terms of switching time-instants for pulses in the $p^+$ group. In this way, specific constraints on the switching time-instants for pulses in the $p^+$ group are also established.  
	
	When $l$ is odd, (see Fig. \ref{oddpulse}), the rising-edge of the pulse in $p^+$ group corresponds to the rising-edge of the pulse in the $q^+$ group, the falling-edge of the pulse in the $p^+$ group corresponds to the rising-edge of the pulse in the $r^+$ group, while falling-edges of the pulses in the $q^+$ and $r^+$ groups are related.  Thus, the time-instant $t_{2P+2l -1}$ in the $q^+$ group is related to $t_{2l-1}$ in the $p^+$ group as follows (see Fig. \ref{oddpulse}):  
	\begin{equation} t_{2l-1} = t_{2P+2l-1} - \frac{T}{6},   \end{equation} 
which leads to
	\begin{equation}
		 t_{2P+2l-1} = t_{2l-1} + \frac{T}{6},  \text{ when } l \text{ is odd.} \label{tpqo1}
	\end{equation}
	Similarly, the switching time-instant $t_{2P + 2l}$ in the $q^+$ group is related to  $t_{4P + 2l }$ in the $r^+$ group as:
	\begin{equation} t_{2P + 2l} - \frac{T}{6} = t_{4P+2l} - \frac{T}{3}. \label{trq2} \end{equation}
	But, using formula (\ref{tpr12}), we can replace $t_{4P+2l}$ by $\frac{T}{2} - t_{2P -(2l-1)}$ , where $t_{2P - (2l-1)}$ is a time-instant in the $p^+$ group. Thus, the above equation is reduced to:
\begin{equation}t_{2P + 2l}  =  \frac{T}{3} - t_{2P-(2l-1)}, \text{ when } l \text{ is odd}. \label{tpqo2} \end{equation}
	Equations (\ref{tpqo1}) and (\ref{tpqo2}) relate time-instants in $q^+$ group to time-instants in the $p^+$ group when $l$ is odd. 
	
	From Fig. \ref{oddpulse}, it is also clear that when $l$ is odd,, time $t_{2l}$ in the $p^+$ group and $t_{4P + 2l - 1}$ in the $r^+$ group are related. Indeed, from Fig. \ref{oddpulse} and using formula (\ref{tpr11}), we can derive:
	\begin{equation} t_{2l} = t_{4P+2l-1} - \frac{T}{3} =  \frac{T}{6} - t_{2P -(2l-2)}, \end{equation} 
	which leads to 
	\begin{equation}
		t_{2l} + t_{2P -(2l-2)} = \frac{T}{6},  \text{ when } l \text{ is odd.}  \label{tppo}
	\end{equation}
	Interestingly, both the switching time-instants $t_{2l}$ and $t_{2P -(2l-2)}$ in equation (\ref{tppo}) belong to the $p^+$ group. This reveals that not all switching time-instants in the $p^+$ group are completely independent. Instead, there exist among them mutual algebraic relations of the form (\ref{tppo}). A similar  relation also holds when $l$ is even. This means that the number of independent variables involved in the optimization of the function $E_2$ needs to be performed is considerably smaller than $2N = 6P$. 

Proceeding in the same way as before, the following equations can be derived  when $l$ is even by using Fig. \ref{evenpulse}:

\begin{align} t_{2P + 2l - 1} &= \frac{T}{3} - t_{2P-(2l-2)}, \label{tqpe1} \\  t_{2P+2l} &= t_{2l} + \frac{T}{6},   \label{tqpe2} \\ t_{2l-1} + t_{2P-(2l-1)} &= \frac{T}{6}.  \label{tppe}\end{align} 

	To summarize, we have established that equations (\ref{tpr11}), (\ref{tpr12}), (\ref{tpqo1}), (\ref{tpqo2}), (\ref{tppo}), (\ref{tqpe1}), (\ref{tqpe2}) and (\ref{tppe}) specify the algebraic relations that the switching time-instants $t_1$,..., $t_{2P}$, $t_{2P+1}$,...,$t_{4P}$,  $t_{4P+1}$, ... ,$t_{6P}$ must satisfy to represent three-phase PWM line-voltages with symmetries (S1)-(S3) under the constraints (C1) and (C2). Imposing these relations as equality constraints on the optimization, a symmetry-preserving time-domain PWM optimization technique can be developed. This matter is further discussed in the next section. 
	
	\section{SYMMETRY-PRESERVING OPTIMAL PWM AND NUMERICAL RESULTS}
	
	In section I, we defined the function $E_2$ in equation (\ref{E2}) and expressed it as a function of switching time-instants $t_1$, $t_2,$...$,t_{2N}$. We discussed how minimizing of $E_2$ leads to the minimization of the harmonic content of the PWM output current [see equation (\ref{ethdreln})].  In Section III, we established that $N=3P$, where $P$ is the number of pulses in each of the $p^+$, $q^+$ and $r^+$  groups. Using the notation introduced in the previous sections, we can write the objective function as $E_2(t_1,..., t_{2P}, t_{2P+1},..., t_{4P}, t_{4P+1}, ..., t_{6P})$. 
	
	We also established that switching time-instants in the $q^+$ and $r^+$  groups can be obtained from switching time-instants in the $p^+$ group, using equations (\ref{tpr11}),(\ref{tpr12}),(\ref{tpqo1}), (\ref{tpqo2}), (\ref{tqpe1}) and (\ref{tqpe2}). Moreover, equations (\ref{tppo}) and (\ref{tppe}) reveal that some switching time-instants in the $p^+$ group are also related. It can be shown that, when $P$ is odd, there are only $\frac{3P-1}{2}$ independent time-instants, in the sense that with a knowledge of these $\frac{3P-1}{2}$  time-instants, all the $6P$ time-instants can be completely determined using equations (\ref{tpr11})-(\ref{tppe}). This dramatic reduction in the number of independent variables over which the optimization is performed $\Big( \text{from }6P \text{ to } \frac{3P-1}{2} \Big)$ greatly simplifies the numerical computation of the optimization problem.
	
	It is apparent that the time-instants $ t_1, t_2,..., t_{6P} $ must be strictly monotonically increasing. This constraint can be expressed as the following (non-strict) inequality constraints: 
	\begin{flalign} 
	 &&t_{j+1} &\geq t_j + \tau  > t_j, \text{ for all } j = 1, ... , 6P - 1, &  \label{ineq11} \\ 
	&\text{and } & \hfill \hfill  \frac{T}{2} &\geq t_{6P} + \tau > t_{6P}, \text{ where } \tau > 0. &  \label{ineq22}
	\end{flalign}
	The inequality constraints in (\ref{ineq11}) and (\ref{ineq22}) are used to numerically implement \cite{opt1,opt2,opt3} the strict monotonicity condition since most numerical optimization solvers do not accept strict inequalities as inputs. It is worthwhile to mention that, if we define:
	\begin{equation} 
		\Delta t_i = t_{i}-t_{i-1}, \text{   for all    } i = 1, ... , 6P+1, 
	\end{equation} then, the strict monotonicity constraint can be expressed as:
	\begin{equation}
		\Delta t_i > 0, \text{   for all    } i = 1, ... , 6P+1.
	\end{equation} 
	The latter inequalities define a convex region (cone) \cite{opt1}. For this reason, it may be advantageous to use the variable $\Delta t_i$ for numerical minimization.	 
	
	It is apparent that the optimal PWM depends on the parameters $R$, $L$ and $T$. It turns out that this dependence can be expressed in terms of a function of only \textit{one} dimensionless parameter. Indeed, this can be accomplished by introducing the following dimensionless parameter:
	\begin{flalign} 
	& &\alpha &=  \frac{RT}{L}, & & \label{alpha} \end{flalign}
	and using the scaled-time:
	\begin{equation}
		\beta = \frac{t}{T}, \quad \quad \Big( \beta_j = \frac{t_j}{T}, \text{ for } \text{all }  j = 1, 2, ... , 2N \Big)
		\label{beta}
	\end{equation}
	as well as voltages:
	\begin{equation}
		v_{R,ab}(t) = Ri_{ab}(t),
		\label{vrab}
	\end{equation}
	\begin{equation}
		\tilde{v}_{R,ab}(\beta) = v_{R,ab}(\beta T),
		\label{tildevrab}
	\end{equation}
	\begin{equation}
	v_{R,a}(t) = R i_{a}(t),
	\label{vRa}
	\end{equation}
	\begin{equation}
	\tilde{v}_{R,a}(\beta) = v_{R,a}(\beta T),
	\end{equation}
	and coefficients 
	\begin{equation}
		B_j = R A_j, \quad \text{ for all } j = 1,2,..., 2N+1.
		\label{bjdef}
	\end{equation}
	Now, equations (\ref{it}),  (\ref{A2k1}), (\ref{A1}) and (\ref{Aj}), can be rewritten as follows
	\begin{equation}
	\tilde{v}_{R,ab}(\beta)=\begin{cases}
	B_{2j+1}e^{-\alpha \beta}, & \text{if $ \beta_{2j} < \beta < \beta_{2j+1} $},\\
	B_{2j+2}e^{- \alpha \beta} + V_0, & \text{if $ \beta_{2j+1} < \beta < \beta_{2j+2}$},
	\end{cases} \label{tvrabb}
	\end{equation}
	where,
	\begin{equation} B_{2N+1} = V_0 \frac {\sum_{j=1}^{2N}{(-1)^j e^{\alpha \beta_j}}}{1 + e^{-\frac{\alpha}{2}}},   \label{B2k1} \end{equation}
	\begin{equation} B_1 = -V_0 \frac {\sum_{j=1}^{2N}{(-1)^j e^{\alpha \beta_j}}}{1 + e^{-\frac{\alpha}{2}}} e^{-\frac{\alpha}{2}} , \label{B1} \end{equation}  
	and
	\begin{equation} 
	B_j = B_1 + V_0 \sum_{n=1}^{j-1}{(-1)^n e^{\alpha \beta_n }} \; \;  \text{ for } j=2,3,...,2N. \label{Bj} 
	\end{equation}
	It is evident that (since $N=3P$):
	\begin{equation}
		\tilde{v}_{R,ab}(\beta) = \tilde{v}_{R,ab}(\beta,  \beta_1,...,\beta_{6P}). \label{funcb} 
	\end{equation}
	Thus, using the formula (\ref{vRa}) along with equation (\ref{ia})  and equations (\ref{vrab})-(\ref{tildevrab}), we obtain:
	\begin{equation}
	\begin{split}
	\tilde{v}_{R,a}(\beta, \beta_1,.. &,\beta_{6P}) =\frac{1}{3}\bigg[ 	\tilde{v}_{R,ab}(\beta,  \beta_1,...,\beta_{6P}) \nonumber \\ &- 	\tilde{v}_{R,ab}\bigg(\beta + \frac{1}{3} ,  \beta_1,...,\beta_{6P} \bigg)\bigg].
	\end{split}
	\end{equation}
	 It is clear from the above equation that  $\tilde{v}_{R,a}(\beta, \beta_1,...,\beta_{6P})$ depends only on the parameter $\alpha$. 
	Similarly, the desired fundamental component of the output voltage $v_{R,a}(t)$ can be expressed as:
	  \begin{equation}
	 \tilde{v}_{R,a,1}(\beta) = v_{R,a,1}(\beta T) = V_{R,m} \sin \bigg(2\pi\beta - \phi - \frac{\pi}{6}\bigg) ,
	 \label{va1}
	 \end{equation}
	 where
	 \begin{equation} V_{R,m} =  \frac{V_m}{\sqrt{3} \sqrt{{1 + \frac{4\pi^2}{\alpha^2}}}}, \label{VRm} \end{equation}
	 and 
	 \begin{equation}
	 	\tan \phi = \frac{2\pi}{\alpha}. 
	 \end{equation}
	 If we define the function: 
	 	\begin{equation}
	 	\begin{split}
	 	\tilde{E}_{2}(\beta_1, ... , &\beta_{6P})=\int_{0}^{\frac{1}{2}} \bigg[ \tilde{v}_{R,a}(\beta,\beta_1,..., \beta_{6P}) \nonumber \\ &- V_{R,m} \sin \bigg(2\pi\beta - \phi - \frac{\pi}{6}\bigg)\bigg]^2 d\beta,  \label{Eb2}
	 	\end{split} 
	 	 \end{equation}
	  it can be easily verified that 
	  \begin{equation}
	  	 \tilde{E}_2(\beta_1, ..,\beta_{6P}) = \frac{R^2}{T}E_2(t_1, ..,t_{6P}). \label{Etilde2}
	  \end{equation}
	 
	 Hence, using the above time-scaling, the effects of the parameters $R$, $L$ and $T$ on the solutions to the optimization problem can be accounted for by using only the parameter $\alpha$.  
	
  Now, the current-harmonics optimization problem in the standard form \cite{opt1,opt2,opt3} can be stated as follows. 
	
	Minimize the function: $\tilde{E}_2(\beta_1, \beta_{2},... )$ as defined in (\ref{Etilde2}), subject to the strict-monotonicity constraints defined in (\ref{ineq11})-(\ref{ineq22}), as well as constraints (\ref{fund})  and (\ref{shecons}) expressed in terms of the $\beta_j$ variables.
	
	This is the standard problem for constrained non-linear optimization that can be numerically solved using techniques such as Interior Point Methods and Sequential Quadratic Programming \cite{opt2,opt3}.
	
	\begin{figure}[h!]
		\centering
		\includegraphics[width=4in]{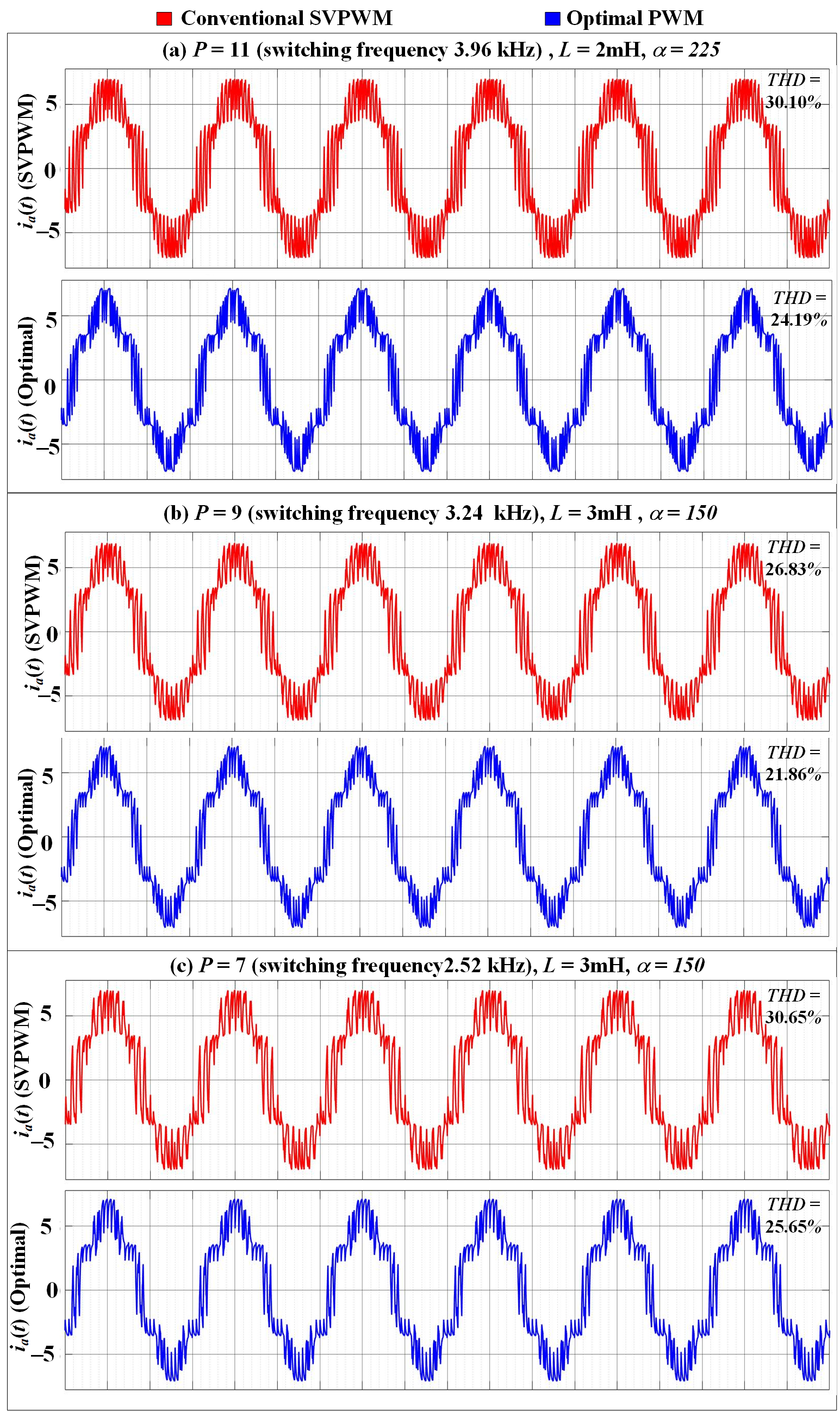}
		\setlength{\belowcaptionskip}{-8pt} \caption{Comparison of conventional SVPWM with Optimal PWM for $I_{m}= 5 A$ and $R = 27 \Omega$ and different values of $P$ and $L$.}
		\label{result1}
	\end{figure}

	Below, some sample calculations performed by using the mentioned techniques are presented.
	These calculations have been performed in MATLAB. Interior-Point Method and Sequential Quadratic Programming methods have been used for optimization. In these calculations, the value of the bus voltage $V_0$ has been taken to be 300 V, and the desired frequency has been chosen to be 60 Hz. The optimization has been performed for different values of inductance $L$ and load resistance $R$ (i.e. for different values of $\alpha$) as well as for various number of pulses $P$. Note that $P$ is related to the switching frequency $f_{sw}$ via the following relation:
	\begin{equation}
	f_{sw} = 6Pf
	\label{fsw}
	\end{equation}
	where $f = \frac{\omega}{2\pi}$. The initial guess for the switching time-instants has been computed according to the conventional Space Vector PWM (SVPWM). The comparative results of the performed calculations are presented in Fig. \ref{result1} and Table 1. 
	
	\begin{table}[h]
	\centering
	\caption{Improvement in THD after optimization when  $I_{m}= 5 A$ and $R = 27 \Omega$}
	\label{table}
	\setlength{ \tabcolsep}{3pt}
	\begin{tabular}{|c|c|c|c|c|c|}
		\hline
		\multirow{2}{*}{$P$} &$f_{sw}$ & L  & \multicolumn{2}{|c|}{THD (in \%)} & \multirow{2}{*}{\% improvement} \\ \cline{4-5}
		&(in kHz)&(in mH)  & SVPWM & Optimal PWM & \\ \hline
		5 & 1.80 & 5 & 28.04 & 23.52 & 16.12\% \\ \hline
		7 & 2.52 & 3  & 30.65 & 25.65 & 16.17\% \\ \hline
		9 & 3.24 & 3  & 26.83 & 21.86 & 18.52\% \\ \hline
		11 & 3.96 & 2  & 30.10 & 24.19 & 19.63\% \\ \hline
	\end{tabular}
	\label{tab1}
\end{table}
\bigskip

\begin{figure}
		\captionsetup[sub]{font=small}
		\centering
		\begin{subfigure}{0.4\textwidth} 
			\includegraphics[width=\linewidth, scale=0.5]{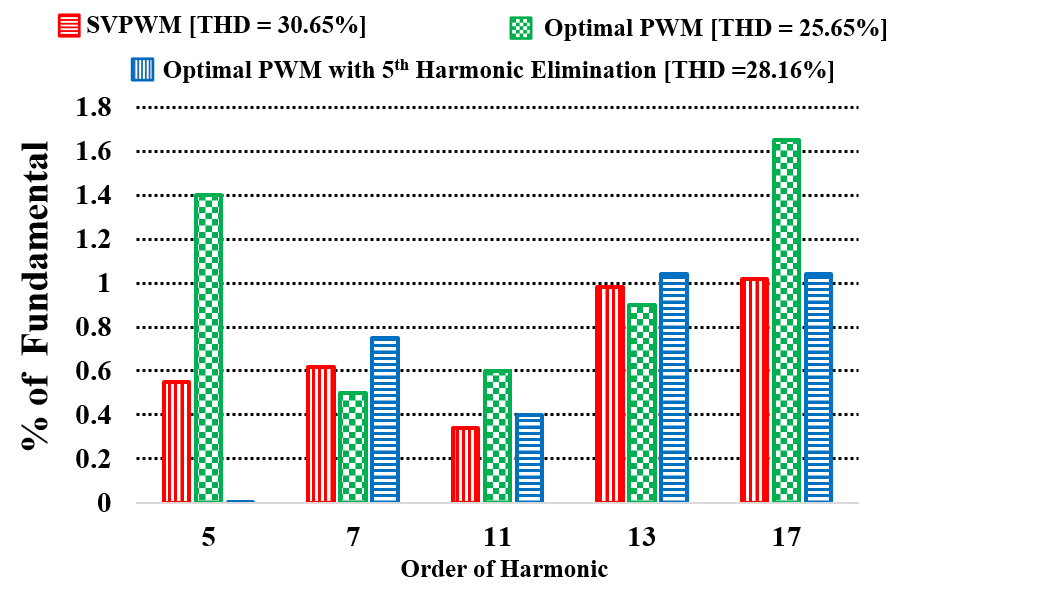}
			\caption{$P = 7, L = 3$ mH [$\alpha = 150$]} 
		\end{subfigure}
		\begin{subfigure}{0.4\textwidth} 
			\includegraphics[width=\linewidth, scale = 0.5]{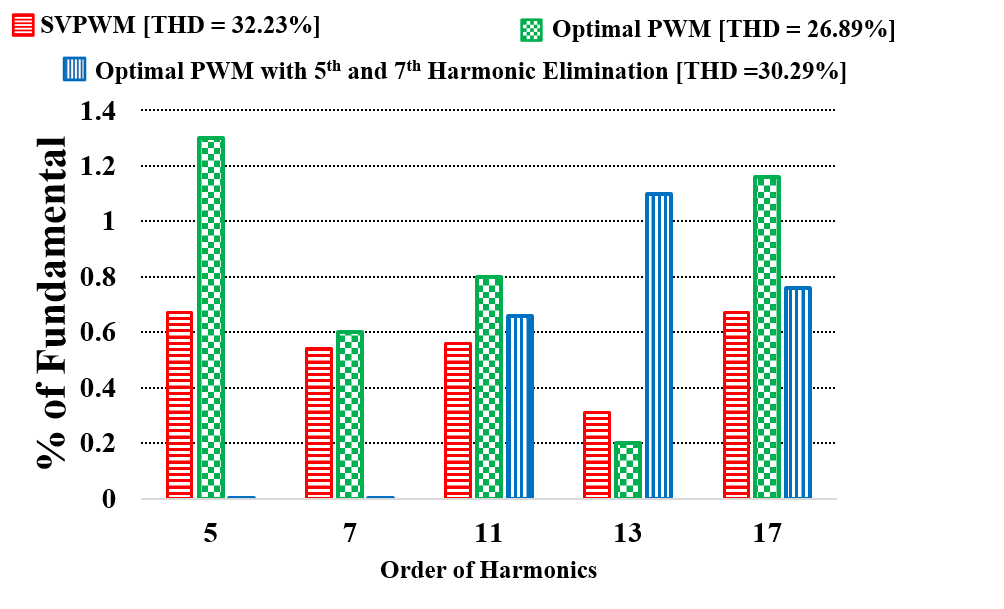}
			\caption{$P = 9, L = 1$ mH [$\alpha = 450$]} 
		\end{subfigure}
		\begin{subfigure}{0.4\textwidth} 
			\includegraphics[width=\linewidth, scale=0.5]{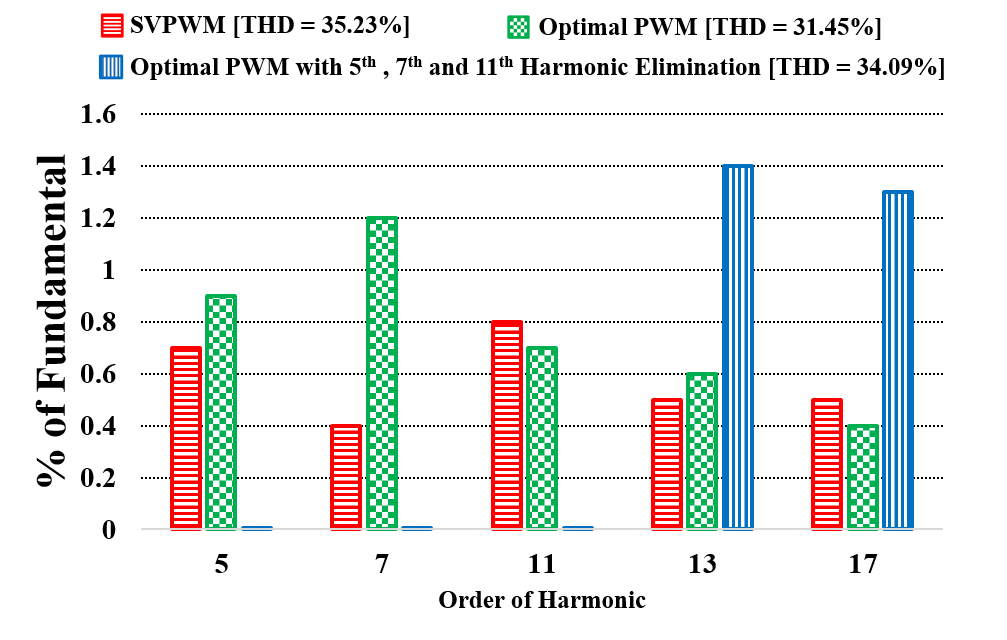}
			\caption{$P = 7, L = 1$ mH [$\alpha = 450$]} 
		\end{subfigure}
		\caption{Computed lower order harmonics  for SVPWM, optimal PWM and optimization with the elimination of (a) $5^{\text{th}}$ order harmonic,  (b) $5^{\text{th}}$ and $7^{\text{th}}$  order harmonics (c) $5^{\text{th}}$, $7^{\text{th}}$ and $11^{\text{th}}$ order harmonic for for  $I_{m}= 5 A$ and $R = 27 \Omega$. Values of parameters $P, L \text{ and } \alpha$ are as specified. Computed THD values are also reported.}  
		\label{result2}
	\end{figure}

	\noindent
	Next, we report the computational results on PWM optimization with elimination of specific lower order harmonics. A major advantage of the time-domain technique is that once the switching time-instants defining PWM voltages are known, exact amplitudes of lower-order harmonics can be computed without any approximation. Some of these computations for conventional SVPWM and optimized PWM are shown in Fig. \ref{result2}. From this figure, it can be seen that in some cases optimization of total harmonic content of the PWM current may result in a slight increase in the percentage of lower-order harmonics. This can be resolved by imposing additional nonlinear constraints of the form (\ref{shecons}) on the optimization such that specific lower-order harmonics can be eliminated.	 It is apparent from this figure that imposing SHE constraints yields sub-optimal performance, as far as THD in the current is concerned. However, even after imposing SHE constraints, better performance than conventional SVPWM, is still achieved in terms of THD.
	
	\section{CONCLUSION}
	A per-phase analysis of three-phase inverters is developed and time-domain analytical expressions are derived for the phase-currents in terms of switching time-instants that describe three-phase PWM voltages. Using these analytical expressions, minimization of harmonics in the output currents and voltages is posed as a standard optimization problem. The use of constrained optimization is proposed for selective harmonic elimination. Furthermore, it is demonstrated that three-phase voltage symmetries, KVL and switching patterns impose specific algebraic constraints on switching time-instants of three-phase PWM voltages. This leads to a significant reduction in the number of independent variables over which the optimization is performed. It is worthwhile to stress that the obtained symmetry constraints on switching-time instants of three-phase PWM voltages are of general nature. These constraints can be essential in the design of different PWM techniques. In the final section of the paper, it is demonstrated that  dependence of the optimized PWM  on parameters $R, L \text{ and } T$ can be expressed in terms of a function of only one dimensionless parameter $\alpha$  by appropriate time-scaling. The numerical results revealing improvements in the harmonic performance of inverters using the optimal time-domain optimization technique are presented. The impact of the optimization on lower-order harmonics is analyzed, and elimination of specific lower-order harmonics using constrained optimization is demonstrated. 
	
	\clearpage

\end{document}